\documentclass[aps,prb,showpacs,twocolumn,superscriptaddress,floats]{revtex4}
\usepackage{amssymb}
\usepackage{amsbsy}
\usepackage{amsmath}
\usepackage{epsfig}

\begin{document}

\title {Superfluid-Insulator transitions of bosons on Kagome lattice at non-integer fillings}

\author{K. Sengupta}
\affiliation{Theoretical Condensed Matter Physics Division,
Saha Institute of Nuclear Physics, 1/AF Bidhannagar, Kolkata-700064, India}
\author{S. V. Isakov}
\affiliation{Department of Physics, University of Toronto, Toronto,
Ontario, Canada M5S 1A7}
\author{Yong Baek Kim}
\affiliation{Department of Physics, University of Toronto, Toronto,
Ontario, Canada M5S 1A7}
\affiliation{School of Physics, Korea Institute for Advanced Study,
Seoul 130-722, Korea}

\date{\today}

\begin{abstract}

We study the superfluid-insulator transitions of bosons on the Kagome
lattice at incommensurate filling factors $f=1/2$ and $2/3$ using a
duality analysis. We find that at $f=1/2$ the bosons will always be
in a superfluid phase and demonstrate that the $T_3$ symmetry of the
dual (dice) lattice, which results in dynamic localization of
vortices due to the Aharonov-Bohm caging effect, is at the heart of
this phenomenon. In contrast, for $f=2/3$, we find that the bosons
exhibit a quantum phase transition between superfluid and
translational symmetry broken Mott insulating phases. We discuss the
possible broken symmetries of the Mott phase and elaborate the theory
of such a transition. Finally we map the boson system to a XXZ spin
model in a magnetic field and discuss the properties of this spin model
using the obtained results.

\end{abstract}


\maketitle

\section {Introduction}

The superfluid (SF) to Mott insulators (MI) transitions of strongly
correlated lattice bosons systems, described by extended Hubbard
models, in two spatial dimensions have recently received a great
deal of theoretical interest. One of the reasons for this renewed
attention is the possibility of experimental realization of such
models using cold atoms trapped in optical lattices \cite{bloch1,
kasevitch1}. However, such transitions are also of interest from a
purely theoretical point of view, since they provide us with a test
bed for exploring the recently developed theoretical paradigm of
non-Landau-Ginzburg-Wilson (LGW) phase transitions \cite{senthil1}. In
the particular context of lattice bosons in two spatial dimensions,
a general framework for such non-LGW transitions has been developed
and applied to the case of square lattice \cite{balents1}. An
application of these ideas for triangular lattice has also been
carried out \cite{burkov1}.

The typical paradigm of non-LGW transitions that has been proposed
in the context of lattice bosons in Refs.\ \onlinecite{balents1} and
\onlinecite{burkov1} is the following. For non-integer rational
fillings $f=p/q$ of bosons per unit cell of the underlying lattice
($p$ and $q$ are integers), the theory of phase transition from the
superfluid to the Mott insulator state is described in terms of the
vortices which are non-local topological excitations of the
superfluid phase, living on the dual lattice.\cite{fisher1, tesanovic1}
These vortices are not
the order parameters of either superfluid or Mott insulating phases
in the usual LGW sense. Thus the theory of the above mentioned phase
transitions are not described in terms of the order parameters on
either side of the transition which is in contrast with the usual
LGW paradigm of phase transitions. Also, as explicitly demonstrated
in Ref.\ \onlinecite{balents1}, although these vortices are
excitations of a featureless superfluid phase, they exhibit a
quantum order which depends on the filling fraction $f$. It is shown
that the vortex fields describing the transition form multiplets
transforming under projective symmetry group (projective
representations of the space group of the underlying lattice). It is
found that this property of the vortices naturally and necessarily
predicts broken translational symmetry of the Mott phase, where the
vortices condense. Since this translational symmetry breaking is
dependent on the symmetry group of the underlying lattice, geometry
of the lattice naturally plays a key role in determining the
competing ordered states of the Mott phase and in the theory of
quantum phase transition between the Mott and the superfluid phases.

In this work, we apply the theoretical framework developed in Refs.\
\onlinecite{balents1} to bosons on Kagome lattice  described by the
extended Bose-Hubbard Hamiltonian
\begin{eqnarray}
H_{\rm boson}&=& -t\sum_{\left<ij\right> }\left( b_i^{\dagger} b_j +
{\rm h.c.} \right) +\frac{U}{2} \sum_i n_i\left(n_i-1\right) \nonumber\\
&& + V \sum_{\left<ij\right>} n_i n_j  -\mu \sum_i n_i
\label{boson1}
\end{eqnarray}
at boson fillings $f=1/2$ and $2/3$. Here $t$ is the boson hopping
amplitude between nearest neighbor sites, $U$ is the on-site
interaction, $V$ denotes the strength of the nearest neighbor
interaction between the bosons and $\mu$ is the chemical potential.

The main motivation of this study is two fold. First, since the
geometry of the underlying lattice plays a significant role in
determining the nature of the Mott phase, we expect that such phases
in Kagome lattice would be distinct from their square
\cite{balents1} or triangular \cite{burkov1} counterparts studied so
far. In particular, the dual of the Kagome lattice is the dice
lattice which is known to have $T_3$ symmetry \cite{vidal1}. It is
well known that the particles in a magnetic field on such lattice
experience a destructive Aharonov-Bohm interference effect at
special values of the external magnetic flux leading to dynamic
localization of the particles. This phenomenon is termed as
Aharonov-Bohm caging in Ref.\ \onlinecite{vidal1}. In the problem at
hand, the vortices reside on the dual (dice) lattice and the boson
filling $f$ acts as the effective magnetic flux for these vortices.
Consequently, we find that at filling $f=1/2$, the vortices become
localized within the Aharonov-Bohm cages \cite{vidal1} and can never
condense. As a result, the bosons always have a featureless
superfluid ground state. Such localization of vortices and
consequently the absence of a Mott phase for bosons is a direct
consequence of the geometry of the Kagome (or it's dual dice)
lattice and is distinctly different from expected and previously
studied behaviors of bosons on square or triangular lattices
\cite{balents1,burkov1} whose dual lattices do not have $T_3$
symmetry. In contrast, for $f=2/3$, we find that there is a
translational symmetry broken Mott phase and discuss the possible
competing ordered states in the Mott phase based on the vortex
theory at a mean-field level. We also address the question of
quantum phase transition from such an ordered state to the
superfluid and write down an effective vortex field theory for
describing such a transition.

The second motivation for undertaking such a study comes from the
interest in physics of XXZ models with ferromagnetic $J_x$ and
antiferromagnetic $J_z$ interaction in a longitudinal magnetic field
$B_l$
\begin{eqnarray}
H_{\rm XXZ} &=& -J_x \sum_{\left<ij\right>} \left( S_i^x S_j^x +
S_i^y
S_j^y \right) + J_z \sum_{\left<ij\right>} S_i^z S_j^z \nonumber\\
&& -B_l \sum_i S^z_i \label{spinmodel1}
\end{eqnarray}
where $J_x >0 $ and $J_z> 0$ are the strengths of transverse and
longitudinal nearest neighbor interactions. Such spin models on
Kagome lattice have been widely studied numerically
\cite{cabra1,sergei1}. Further, the large $J_z$ limit of this model
(in the presence of an additional transverse magnetic field) has
also been studied before \cite{sondhi1}. A couple of qualitative
points emerge from these studies. First, in the absence of external
field $B_l$, such models on Kagome lattice do not exhibit $S_z$
ordering for any values of $J_z/J_x$. This absence of ordering is an
unique property of the Kagome lattice and it has been conjectured
that the ground state is quantum disordered.\cite{sondhi1}
Second, for $B_l \ne 0$ and net magnetizations
$m=\left<S_z\right>=\hbar/6, \hbar/3$, the model exhibit a quantum
phase transition between a featureless state with $\left <S_x
\right> \ne 0$ to a translational symmetry broken ordered state with
finite $m$. Using a simple Holstein-Primakoff transformation which
maps the $H_{\rm XXZ}$ (Eq.\ \ref{spinmodel1}) to hardcore
Bose-Hubbard model $H_{1}$ (Eq.\ \ref{hardcore}), we show that it is
possible to understand both of these features analytically at least
at a qualitative level. The absence $S_z$ ordering for $B_l=0$
(which corresponds to average boson filling $f=1/2$) turns to be a
natural consequence of Aharonov-Bohm caging phenomenon discussed
earlier. Further, the results from the analysis of the Boson model
at $f=2/3$ can also be carried over to study the possible $S_z$
orderings of the spin model at net magnetization $\hbar/3$ which
allows us to make contact with recent numerical studies in Refs.\
\onlinecite{sergei1} and \onlinecite{cabra1}.

The organization of the paper is as follows. In the next section, we
map between the spin model $H_{\rm XXZ}$ to an boson model and carry
out a duality analysis of this boson model. The dual Lagrangian so
obtained is analyzed in Sec.\ \ref{phases} for both filling factors
$f=1/2$ and $2/3$. This is followed by a discussion of the results
in Sec.\ \ref{conclusion}. Some details of the calculations are
presented in Appendices\ \ref{appc}, \ref{appa} and \ref{appb}.

\section{Duality Analysis}
\label{dual1}

To analyze the spin model $H_{\rm XXZ}$, our main strategy is to map
it to a Boson model using the well-known Holstein-Primakoff
transformation
\begin{eqnarray}
S^+_i = (S_i^-)^{\ast} \equiv b_i \quad S_i^z \equiv
\left(b_i^{\dagger} b_i  - \frac{1}{2} \right)
\end{eqnarray}
Such a transformation maps $H_{\rm XXZ}$ to $H_{1}$ given by
\begin{eqnarray}
 H_1 &=& -t\sum_{\left<ij\right> }\left( b_i^{\dagger} b_j +
{\rm h.c.} \right) \nonumber\\
&& + V \sum_{\left<ij\right>} (n_i-f) (n_j-f)
 \label{hardcore}
\end{eqnarray}
with the hardcore constraint $n_i \le 1$ on each site. The
parameters of the hardcore boson model $H_1$ are related to those of
$H_{\rm XXZ}$ as
\begin{eqnarray}
J_{x}&=& 2t, \quad J_z= V, \quad B_l = zV\left(f-\frac{1}{2}\right)
\label{coeff}
\end{eqnarray}
where $z$ is the boson coordination number.
In the Mott phases, the average boson density is locked to some
number and this will be $f$ in Eqs.\ref{hardcore} and \ref{coeff}.
In this work,
we will consider the superfluid-insulator transitions
with the average boson density fixed across the transition.

In what follows, we would carry out a duality analysis of the
hardcore boson model. Such a duality analysis of the boson model is
most easily done by imposing the hardcore constraint in $H_1$ (Eq.\
\ref{hardcore}) by a strong on-site Hubbard term potential to obtain
$H_{\rm boson}$ (Eq.\ \ref{boson1}). In the limit of strong U the
qualitative nature of the phases of the model is expected to be the
same as that of the hardcore model. Hence for rest of the work, we
shall consider the boson model $H_{\rm boson}$. Also, since we are
going to carry out a duality analysis of this model, we shall not
bother about precise relations between parameters of $H_1$ and
$H_{\rm boson}$, but merely represent $\mu$ to be a chemical
potential which forces a fixed filling fraction $f$ of bosons as
shown in previous work \cite{balents1}.

\begin{figure}
\rotatebox{0}{
\includegraphics[width=0.9\linewidth]{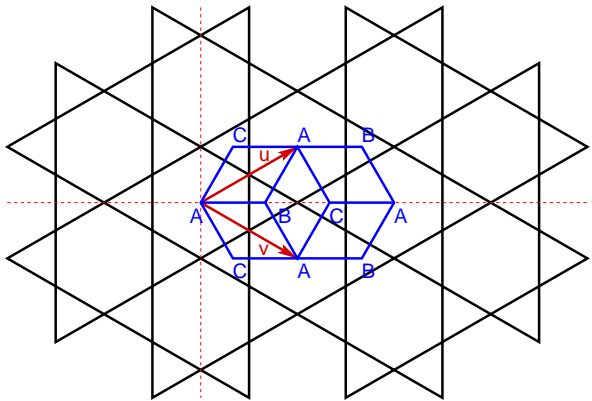}}
\caption{Kagome lattice (thin black line) and the unit cell of its
dual dice (thick blue lines) lattice. The three inequivalent sites
$A$, $B$ and $C$ of the dice lattice are shown in the figure. The
sides of the rhombus $ABAC$ have length $a$ and angles $\pi/3$. The
basis vectors ${\bf u} =(3a/2,\sqrt{3}a/2)$ and ${\bf v}
=(3a/2,-\sqrt{3}a/2)$ that span the dice lattice are shown as
arrowed lines. Note that the $A$ sides reside in the center of the
hexagons of the Kagome lattice while the $B$ and $C$ sites reside in
the center of the triangles. } \label{fig01}
\end{figure}

The dual representation of $H_{\rm boson}$ can be obtained in the
same way as in Refs. \onlinecite{balents1,burkov1}. The details of
the duality transformation are briefly sketched in Appendix\
\ref{appc}. The dual theory turns out to be a theory of U(1)
vortices, residing on the sites of the dual (dice) lattice shown in
Fig.\ \ref{fig01}, coupled to a fictitious magnetic field which
depends on the boson filling $f$. The dual action is given by
\begin{eqnarray}
S_{d} &=& \frac{1}{2e^2} \sum_{p} \left( \epsilon_{\mu \nu \lambda}
\Delta_{\nu} A_{b\lambda} - f \delta_{\mu \tau} \right)^2 \nonumber\\
&& -y_v \sum_{b,\mu} \left(\psi_{b+\mu}^{\dagger} e^{2\pi i A_{b
\mu}} \psi_b
+{\rm h.c} \right) \nonumber\\
&& + \sum_b  r |\psi_b|^2 + u |\psi_b|^4 + ....
 \label{dual}
\end{eqnarray}
where $\psi_b$ are the vortex field living on the site $b$ of the
dual (dice) lattice, $A_{b\mu}$ is the U(1) dual gauge field so that
$\epsilon_{\tau \nu \lambda} \Delta_{\nu} A_{b \lambda}= n_i$ where
$n_i$ is the physical boson density at site $i$, $\sum_p$ denotes
sum over elementary rhombus of the dice lattice$, \Delta_{\mu}$
denotes lattice derivative along $\mu=x,y,\tau$, and $f$ is the
average boson density.
The magnetic field seen by the vortices is therefore the
physical boson density. Note that the vortex action $S_d$ is not
self-dual to the boson action obtained from $H_{\rm boson}$.
Therefore we can not, in general, obtain a mapping between the
parameters of the two actions, except for identifying
$\epsilon_{\tau \nu \lambda} \Delta_{\nu} A_{b\lambda}$ as the
physical boson density \cite{balents1,burkov1}. Therefore, in the
remainder of the paper, we shall classify the phases of this action
based on symmetry consideration and within the saddle point
approximation as done in Ref.\ \onlinecite{balents1}.

The transition from a superfluid ($\left<\psi_b \right> =0$) to a
Mott insulating phase in $S_d$ can be obtained by tuning the
parameter $r$. For $r>0$, we are in the superfluid phase. Note that
the saddle point of the gauge fields $A_{b\mu}$ in action
corresponds to $\epsilon_{\tau \nu \lambda} \Delta_{\nu} {\bar
A}_{b\lambda}= f$, so that the magnetic field seen by the vortices
is pinned to the average boson filling $f$. Now as we approach the
phase transition point $r=0$, the fluctuations about this saddle
point ($\left<\psi_b \right> =0$, ${\bar A}_{by} = f x$) increase
and ultimately destabilize the superfluid phase in favor of a Mott
phase with $\left<\psi_b \right> \ne 0$. Clearly, in the above
scenario, the most important fluctuations of the vortex field
$\psi_b$ are the ones which has the lowest energy. This prompts us
to detect the minima of the vortex spectrum by analyzing the kinetic
energy term of the vortices
\begin{eqnarray}
H_{\rm kinetic} &=& -y_v \sum_{b,\alpha } \left(\psi_{b+\alpha}
^{\dagger} e^{2\pi i {\bar A}_{b}} \psi_{b} + {\rm h.c} \right)
\label{kinetic}
\end{eqnarray}
where the sum over $\alpha$ is carried out over the six-fold (for
$A$ sites in Fig.\ \ref{fig01}) or three-fold (for $B$ and $C$
sites) coordinated sites surrounding the site $b$.

The analysis of $H_{\rm kinetic}$ amounts to solving the Hofstadter
problem on the dice lattice which has been carried out in Ref.\
\onlinecite{vidal1}. We shall briefly summarize the relevant part of
that analysis. It is found that the secular equation obtained from
$H_{\rm kinetic}$ for boson filling factor $f$ is given by
\cite{vidal1}
\begin{eqnarray}
\epsilon \psi_A(x,\kappa) &=&  \psi_B(x+a) + \psi_C (x-a) \nonumber\\
&& + 2\cos\left[\frac{\gamma}{a}\left(x-a/4\right) + \kappa \right]
\psi_B(x-a/2) \nonumber\\
&& + 2\cos\left[\frac{\gamma}{a}\left(x+a/4\right) + \kappa \right]
\psi_C(x+a/2) \nonumber\\
\epsilon \psi_B(x,\kappa) &=& \psi_A (x-a)\nonumber\\
&& + 2\cos\left[\frac{\gamma}{a}\left(x+a/4\right) + \kappa \right]
\psi_A(x+a/2) \nonumber\\
\epsilon \psi_C(x,\kappa) &=&  \psi_A (x-a) \nonumber\\
&& + 2\cos\left[\frac{\gamma}{a}\left(x-a/4\right) + \kappa \right]
\psi_A(x-a/2) \label{secu1}
\end{eqnarray}
 where $A$, $B$ and $C$ denote inequivalent sites of the dice
lattice, $\gamma=2\pi f = 2\pi \phi/\phi_0$ for the boson filling
$f$, and $\phi$ is the dual flux through an elementary rhombus of
the lattice. Here $\psi = (\psi_A,\psi_B, \psi_C)$ is the vortex
wavefunction which has been written as $\psi_b \equiv \psi(x,y)=
\psi(x,\kappa) \exp(i\kappa n)$ with  $\kappa = {\sqrt 3}k_y a/2$
and integer $n=2y/(\sqrt{3}a)$, a is the lattice spacing of the dice
lattice, and $\epsilon$ is the energy in units of vortex fugacity
$y_v$. For obtaining the solutions corresponding to $\epsilon \ne
0$, we eliminate for $\psi_B$ and $\psi_C$ from Eq.\ \ref{secu1} to
arrive at an one-dimensional equation for $\psi_A(x=3ma/2,\kappa)
\equiv \psi_A(m,\kappa)$
\begin{eqnarray}
&& \left(\frac{\epsilon^2 -6}{2 \cos\left(\gamma/2\right)} \right)
\psi_A(m,\kappa) =  2\cos\left(3\gamma m +2\kappa \right)
\psi_A(m,\kappa) \nonumber\\
&& +2\cos\left[\frac{3\gamma}{2}\left(m-1/2\right) +\kappa\right]
\psi_A(m-1,\kappa) \nonumber\\
&& +2\cos\left[\frac{3\gamma}{2}\left(m+1/2\right) +\kappa \right]
\psi_A(m+1,\kappa)  \label{secu2}
\end{eqnarray}
where we have used the fact that with our choice of origin the $A$
sites have $x=3ma/2$ with integer $m$. For rational boson filling
$f=p/q$, Eq.\ \ref{secu2} closes after translation by $q$ periods.
However, for $q=3q'\, [q'\in {\rm integer}]$, this periodicity is
reduced to $q'$ \cite{vidal1}.

A key feature of the wavefunction $\psi$ that we would be using in
the subsequent analysis is its transformation property under all
distinct space group operations of the dice lattice. Therefore we
collect all such transformations here for any general filling $f$.
Such operations for the dice lattice, as seen from Fig.\
\ref{fig01}, involve translations $T_u$ and $T_v$ along vectors
$u=(3a/2,{\sqrt 3}a/2)$ and $v=(3a/2,-{\sqrt 3}a/2)$, rotations $\pi
n/3$ for any integer n, and reflection $I_x$ about the $x$ axis. It
is easy to check that these are the basic symmetry operations that
must leave the Hamiltonian invariant and all other symmetry
operations can be generated by their appropriate combinations.
Following methods outlined in Ref.\ \onlinecite{balents1}, one finds
the following transformation properties for the wavefunctions:
\begin{eqnarray}
T_{\alpha=u,v}: \psi(x,y) \rightarrow \psi(x-\alpha_x,y-\alpha_y)
\omega^{\alpha_x y 2/\sqrt{3}} \label{translation} \\
I_x:\psi(x,y) \rightarrow\psi^{\ast}(x,-y)
\label{reflection} \\
R_{\pi/3}: \psi_A(x,y) \rightarrow \psi_A\left([x+\sqrt{3}y
]/2,[y-\sqrt{3}
x ]/2 \right)\nonumber\\
 \times  \omega^{\left[\left(y^2-x^2\right)/4 + \sqrt{3}x
y/2\right]}\nonumber\\
\psi_B(x,y) \rightarrow \psi_C\left([x+\sqrt{3}y ]/2,[y-\sqrt{3}
x ]/2 \right)\nonumber\\
 \times  \omega^{\left[\left(y^2-x^2\right)/4 + \sqrt{3}x
y/2\right]} \nonumber\\
\psi_C(x,y) \rightarrow \psi_B\left([x+\sqrt{3}y ]/2,[y-\sqrt{3}
x ]/2 \right)\nonumber\\
 \times  \omega^{\left[\left(y^2-x^2\right)/4 + \sqrt{3}x
y/2\right]}\label{rot2}
\end{eqnarray}
where $(x,y)$ are the coordinates of the dice lattice and
$\omega = \exp(2 \pi i f)$, Here $\psi$ may denote either
$\psi_A(x=3ma/2,y=\sqrt{3}na/2)$ or $\psi_B$ and $\psi_C$ at
appropriate positions within the unit cell. Note that all other
above mentioned symmetry operations can be generated by combination
of the operations listed in Eqs.\ \ref{translation}..\ref{rot2}. For
example $I_y = I_x R_{\pi} = I_x R_{\pi/3}^3$ and under this
transformation $\psi_A(x,y) \rightarrow \psi_A^{\ast}(-x,y)$ and
$\psi_{B,C}(x,y) \rightarrow \psi_{C,B}^{\ast}(-x,y)$. Some details
of the algebra leading to Eqs.\ \ref{translation}..\ref{rot2} is
sketched in Appendix \ref{appa}.

In what follows, we are going to analyze the vortex theory for
$f=1/2$ and $f=2/3$ which turns out to be the simplest possible
physically interesting fractions to analyze.

\section {Phases for $f=1/2$ and $2/3$}
\label{phases}

\subsection{f=1/2}

The key difference between square and triangular lattices analyzed
previously \cite{balents1, burkov1}and the Kagome lattice studied in
this work, comes out when we study the filling fraction $f=1/2$. From
Eq.\ \ref{secu2}, we find that for $f=1/2$, $\gamma=\pi$, so that
the entire spectrum collapses into three infinitely degenerate bands
$\epsilon= 0,\pm \sqrt{6}$. This situation is in sharp contrast to
the square or the triangular lattices where the vortex spectrum  for
rational boson fillings $f=p/q$ has a fixed number of well defined
minima. It is possible that vortex-vortex interaction may make the
degeneracy finite, but it is unlikely to be lifted completely by
interaction \cite{vidal1}. Therefore the vortex band has zero
or at least extremely small bandwidth with no well-developed minima
at any wavevector which implies that the vortices shall never
condense \cite{comment1}. Consequently, we do not expect to have a
Mott insulating state at $f=1/2$, as seen in previous Monte Carlo
studies $H_1$ on Kagome lattice \cite{sergei1}. Comparing the
parameters of $H_1$ and $H_{\rm XXZ}$ (Eq.\ \ref{coeff}), we find
that such an absence of Mott phase also implies that XXZ models on
Kagome lattice will have no $S_z$ ordering for $B_l=0$. This result
was also obtained by numerical study of the related transverse field
Ising model on the Kagome lattice\cite{sondhi1}.

\begin{figure}
\rotatebox{0}{
\includegraphics[width=0.5\linewidth]{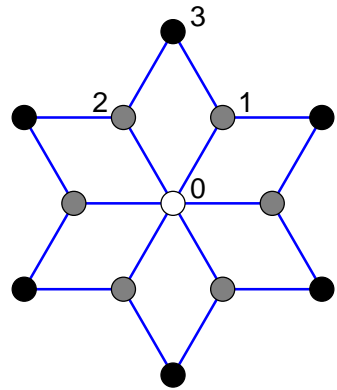}}
\caption{Aharonov-Bohm cage for a vortex on site $0$ of the dice
lattice. The probability amplitude for the particle to be at $3$
vanishes due to destructive interference between amplitudes from the
paths $0\rightarrow 1 \rightarrow 3$ and $0\rightarrow 2 \rightarrow
3$. Consequently the vortex can never propagate beyond the rim of
the cage (black circles) and is dynamically localized. Analogous
cages can be constructed for vortices on sites $B$ and $C$
\cite{vidal1}.} \label{fig02}
\end{figure}

Physically, the collapse of the vortex spectrum into three
degenerate bands can be tied to localization of the vortex within
the so-called Aharonov-Bohm cages as explicitly demonstrated in
Refs.\ \onlinecite{vidal1}. An example of such a cage is shown in
Fig.\ \ref{fig02}. A vortex whose initial wavepacket is localized at
the central (white) $A$ site can never propagate beyond the black
sites which form the border of the cage. This can be understood in
terms of destructive Aharonov-Bohm interference: The vortex has two
paths $0\rightarrow1\rightarrow 3$ and $0\rightarrow 2\rightarrow 3$
to reach the rim site $3$ from the starting site $0$. The amplitudes
from these paths destructively interfere for $f=1/2$ to cancel each
other and hence the vortex can never propagate beyond the rim of the
cage. Similar cages can be constructed for vortices with initial
wavepacket at sites $B$ and $C$ \cite{vidal1}. This dynamic
localization of the vortex wavepackets in real space, demonstrated
and termed as Aharonov-Bohm caging in Ref.\ \onlinecite{vidal1},
naturally leads to spreading of the vortex wavepacket in momentum
space. Hence the vortex spectrum has no well-defined minima at any
momenta leading to the absence of vortex condensation.

The question naturally arises regarding the nature of the boson
ground state at $f=1/2$ (or equivalently ground state of XXZ model
at $B_l=0$ and $J_z \gg J_x$). The dual vortex theory studied here can
only predict the absence of vortex condensation due to dynamic
localization of the vortices leading to persistence of superfluidity
(or $S_x$ ordering for XXZ model) at all parameter values. Beyond
this, numerical studies are required to discern the precise nature
of the state. Previous studies of the transverse Ising model, which is
related to the XXZ model, in Ref.\ \onlinecite{sondhi1} have conjectured
this state to be a spin liquid state. However, the Monte Carlo studies
of the XXZ model\cite{sergei1} did not find any sign of spin liquid
state.

\subsection{f=2/3}

In contrast to $f=1/2$, the physics for $f=2/3$ (or $f=1/3$) turns
out to be quite different. In this case, we do not have any dynamic
localization effect and the vortex spectrum has well defined minima.
Note that for $f=2/3$, Eq.\ \ref{secu2} closes after one period and
the magnetic Brillouin zone becomes identical to the dice lattice
Brillouin zone without a magnetic field \cite{vidal1}.  As a result,
Eq.\ \ref{secu2}, admits a simple analytical solution
\begin{eqnarray}
\epsilon &=& - \left( 6 - 2 \cos\left(2 \kappa \right) + 4
\cos\left(\kappa\right) \cos(k_m) \right)^{1/2} \nonumber\\
\psi_A (x,y) &=& \exp\left(i m 3k_xa/2 + \kappa n\right)
\label{sol1}
\end{eqnarray}
where we have $\kappa \in (0, \pi)$ and $k_x \in (0,4\pi/3)$, and
$k_m=3k_x a/2$. The spectrum has two minima in the magnetic
Brillouin zone at $ (k_x a, \kappa) = (0, \pi/3)\, {\rm and}
\,(2\pi/3, 2 \pi/3)$. The eigenfunctions $\psi = \left(\psi_A,\psi_B
,\psi_C \right)$ , corresponding to these two minima can be obtained
from Eqs.\ \ref{sol1} and \ref{secu1}. They are
\begin{eqnarray}
\psi_1 &=& \exp(i \pi n/3)(c,-c,0)\nonumber\\
\psi_2 &=& \exp \left[ \frac{2\pi i n}{3} + i\pi m\right](c,0,-c)
\label{efunc}
\end{eqnarray}
where $c$ is a normalization constant. Thus the low energy
properties of the vortex system can then be characterized in terms
of the field $\Phi$:
\begin{eqnarray}
\Phi &=& \psi_1({\bf x}) \varphi_1({\bf x},t) +\psi_2({\bf x})
\varphi_2({\bf x},t) \label{vfunc}
\end{eqnarray}
where $\varphi_{1(2)}$ are fields representing low energy
fluctuations about the minima $\psi_{1(2)}$. As shown in Ref.\
\onlinecite{balents1}, the transition from the superfluid to the
Mott phase of the bosons can be understood by constructing a low
energy effective action in terms of the $\varphi$ fields.

To obtain the necessary low energy Landau-Ginzburg theory for the
vortices, we first consider transformation of the vortex fields
$\varphi$ under the symmetries of the dice lattice. Using Eqs.\
\ref{translation},\ref{reflection},\ref{rot2}, \ref{efunc}, and
\ref{vfunc}, one gets
\begin{eqnarray}
T_{u}: \varphi_{1} \rightarrow \varphi_{1} \exp(-i\pi/3)
\quad \varphi_{2} \rightarrow \varphi_{2} \exp(i\pi/3) \nonumber\\
T_{v}: \varphi_{1} \rightarrow \varphi_{1} \exp(i\pi/3)
\quad \varphi_{2} \rightarrow \varphi_{2} \exp(-i\pi/3) \nonumber\\
I_x: \varphi_{1(2)} \rightarrow \varphi^{\ast}_{1(2)}\nonumber\\
I_y: \varphi_{1(2)} \rightarrow \varphi^{\ast}_{2(1)}\nonumber\\
R_{2\pi/3}: \varphi_{1(2)}\rightarrow \varphi_{1(2)}\nonumber\\
R_{\pi/3}: \varphi_{1(2)}\rightarrow \varphi_{2(1)} \label{transv1}
\end{eqnarray}
Some details of the algebra leading to Eq.\ \ref{transv1} is given
in App.\ \ref{appb}.  It is also instructive to write the two fields
$\varphi_1$ and $\varphi_2$ as two components of a spinor field
$\chi = (\varphi_1,\varphi_2)$. The transformation properties of
$\chi$ can then be written as
\begin{eqnarray}
T_{u}: \chi \rightarrow \chi \exp(-i \tau_z \pi/3) \nonumber\\
T_{v}: \chi \rightarrow \chi\exp(i\tau_z \pi/3)\nonumber\\
I_x: \chi\rightarrow \chi ^{\ast}\nonumber\\
I_y: \chi \rightarrow \tau_x \chi^{\ast}\nonumber\\
R_{2\pi/3}: \chi\rightarrow \chi \nonumber\\
R_{\pi/3}: \chi \rightarrow \tau_x \chi \label{transv2}
\end{eqnarray}
where $\tau_{x,y,z}$ are the usual Pauli matrices. Note that
$R_{2\pi/3}$ plays the role of identity here where $R_{\pi/3}$ mixes
the two components of the spinor field $\chi$.

The simplest Landau-Ginzburg theory for the vortex fields which
respects all the symmetries is
\begin{eqnarray}
L_v &=& L_v^{(2)}+ L_v^{(4)}+ L_v^{(6)} \\
L_v^{(2)} &=& \sum_{\alpha=1,2} \left[ \left|\left(i\partial_{\mu} -
e A_{\mu}\right) \varphi_{\alpha} \right|^2 +
r \left|\varphi_{\alpha}\right|^2  \right]\\
L_v^{(4)} &=& u\left(\left|\varphi_1\right|^4 +
\left|\varphi_2\right|^4 \right) + v \left|\varphi_1\right|^2
\left|\varphi_2\right|^2 \\
L_v^{(6)} &=& w \left[ \left(\varphi_1^{\ast} \varphi_2 \right)^3 +
{\rm h.c.} \right]
\end{eqnarray}
The above lagrangian density can also be written in terms of the
$\chi$ fields. In particular the sixth order term becomes $L_v^{6} w
\left[\sum_{a=\pm} \left(\chi^{\ast} \tau_a \chi\right)^3
\right]$. This turns out to be the lowest-order term which does not
commute with $\tau_z$ and breaks the $U(1)$ symmetry associated with
the relative phase of the bosons. A simple power counting shows that
$L_v^{(6)}$ is marginal at tree level. Unfortunately, the
relevance/irrelevance of such a term beyond the tree level, which is
of crucial importance for the issue of deconfinement at the quantum
critical point, is not easy to determine in the present context for
two reasons. First, the standard $\epsilon$ and large $N$ expansions
does not yield reliable results for $2+1$D field theories, at least
for $N=2$ which is the case of present interest \cite{chen1}.
Second, while it is known from numerical studies that such a cubic
anisotropy term is relevant for the single vortex species case (for
XY transitions in $2+1$ D), the situation remains far less from
clear for the two vortex case \cite{ashvin1}. We have not been able
to determine the relevance/irrelevance of this term in the present
work.

If it so turns out that $L_v^{(6)}$ is irrelevant,  the situation
here will be identical to that of bosons on square lattice at
$f=1/2$. The relative phase of the vortices would emerge as a
gapless low energy mode at the critical point. The quantum critical
point would be deconfined and shall be accompanied by boson
fractionalization \cite{balents1}. On the other hand, if $L_v^{(6)}$
is relevant, the relative phase degree of the bosons will always
remain gapped and there will be no deconfinement at the quantum
critical point. Here there are two possibilities depending on the
sign of $u$ and $v$. If $u,v<0$ and $w>0$, the transition may become
weakly first order whereas for $u>0$, it remains second order (but
without any deconfinement). The distinction between these scenarios
can be made by looking at the order of the transition and the
critical exponents in case the transition turns out to be second
order \cite{senthil1}.

\begin{figure}
\rotatebox{0}{
\includegraphics[width=0.9\linewidth]{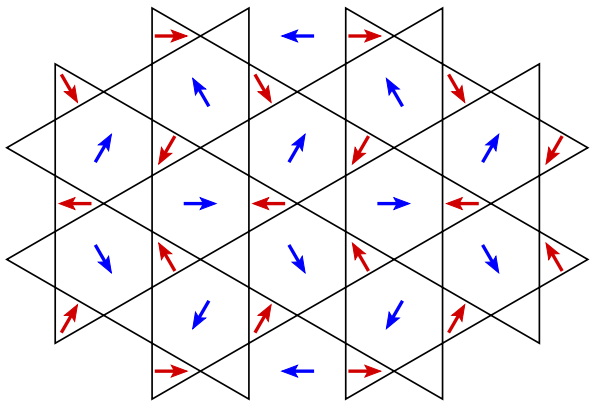}}
\caption{Mean-field vortex state for $v >0$ with $\varphi_1 \ne 0$.
The arrows denote the phase of the vortex wavefunctions $\Phi$ at
the sites of the dual lattice. Note that all the sites of the Kagome
lattice see the same vortex environment and are therefore
equivalent. Hence we do not expect a density wave at the mean-field
level.} \label{fig1}
\end{figure}

\begin{figure}
\rotatebox{0}{
\includegraphics[width=0.9\linewidth]{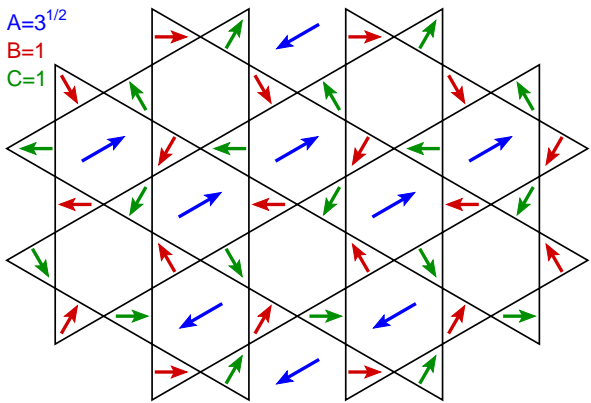}}
\caption{Mean-field vortex state for $v <0$ with $|\varphi_1|
|\varphi_2| =1$ and $\theta_{\rm relative} = \pi/3$. The notations
are same as in Fig.\ \ref{fig1}. The magnitudes of the vortex
wavefunctions are listed on the left, and $A$, $B$ and $C$ refers to
corresponding inequivalent sites of the dual (dice) lattice where
the vortices reside. Here there are six inequivalent sites for the
Kagome lattice which are listed in Fig.\ \ref{fig4}.} \label{fig2}
\end{figure}

The vortices condense for $r<0$ signifying the onset of the Mott
state and now we list the possible Mott phases. Our strategy in
doing so would to be list the sites on the Kagome lattice which sees
identical vortex environments (within mean-field) and then obtain
the possible ground states based on this classification for a fixed
boson filling $f=2/3$. We would like to stress that these results
are obtained within mean-field theory without taking into account
quantum fluctuations.

First let us consider the case $v>0$. Here only one of the vortex
fields condense and consequently the Mott state breaks the symmetry
between $B$ and $C$ sites of the dual (dice) lattice. A plot of
the mean-field vortex wavefunction for $\left<\varphi_1\right> \ne
0$ is shown in Fig.\ \ref{fig1}. The direction of the arrows denote
the phase of the vortex wavefunction at the sites of the dual
lattice while the amplitude of the wavefunctions are listed in the
figures. We find that for this state all the sites of the Kagome
lattice see identical vortex environment and consequently, within
mean-field theory, we do not expect density-wave ordering for the
bosons. However, the up and down triangles have different vortex
environments and one would therefore expect increased effective
kinetic energy of the bosons around either the up or down
triangles. Consequently one expects an equal superposition state of
bosons in which there is equal amplitude of the bosons in the sites
around the up(down) triangles \cite{balents1}.

\begin{figure}
\rotatebox{0} {\includegraphics[width=0.9\linewidth]{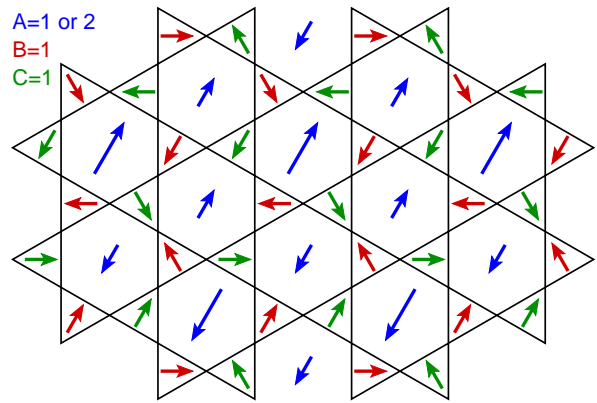}}
\caption{Mean-field vortex state for $v <0$ with $|\varphi_1|
|\varphi_2| =1$ and $\theta_{\rm relative} = 2\pi/3$. This state
also has six inequivalent sites. Notations are same as in Fig.\
\ref{fig2}.} \label{fig3}
\end{figure}

\begin{figure}
\rotatebox{0}{
\includegraphics[width=0.9\linewidth]{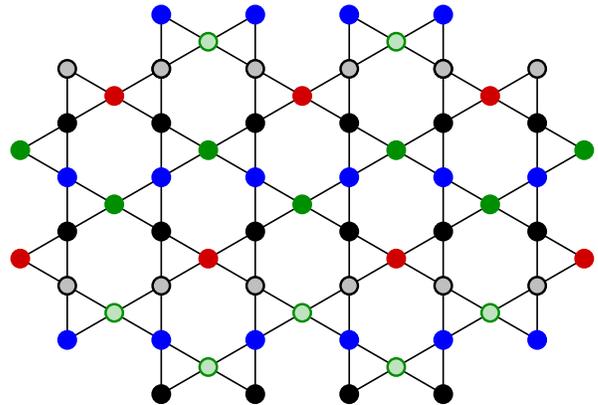}}
\caption{A chart of the inequivalent sites for the mean-field vortex
state for $v <0$ with $|\varphi_1| = |\varphi_2| =1$ and
$\theta_{\rm relative} = \pi/3$ denoted by red, blue, green (open
and closed) and black (open and closed) circles. Inequivalent sites
are defined as those which see a different vortex environment in the
surrounding dual lattice. The relevant vortex environments are
sketched in Figs.\ \ref{fig2} and \ref{fig3}.} \label{fig4}
\end{figure}

\begin{figure}
\rotatebox{0}{
\includegraphics[width=0.9\linewidth]{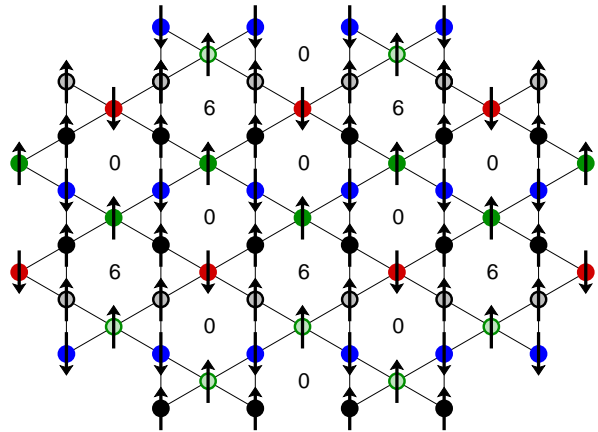}}
\caption{One possible density wave states $\Phi_1$ at $f=2/3$
($m=1/3$ in units of $\hbar/2$) when the black and green (both solid
and empty dots) dots have all filled (or spin up) sites while the
blue and red sites are empty (spin down). The centers of the
hexagons are marked with $6$ and $0$ respectively and shows
the sum of magnetization (in units of $\hbar/2$)of the sites
surrounding the hexagons. This state shows a $3$ by $3$ density wave
pattern as can be seen in the figure. A similar state can be
obtained for $f=1/3$ by replacing the bosons with holes and vice
versa.} \label{fig5}
\end{figure}

For $v<0$, both the vortex fields have non-zero amplitude ${\it
i.e.}$ $|\varphi_1| = |\varphi_2| \ne 0$, and the ground state does
not distinguish between the $B$ and the $C$ sites of the dice
lattice. The relative phase between the vortex fields is fixed by
$L_v^{(6)}$:
\begin{eqnarray}
\theta_{\rm relative} &=& 2\pi p/3 \quad {\rm for}\,\, w<0 \nonumber\\
&=& \pi (2p+1)/3 \quad {\rm for} \,\,w>0 \label{relphase}
\end{eqnarray}
for integer $p$. The plot of the vortex mean-field wavefunctions are
shown in Figs.\ \ref{fig3} for $w>0$ and Fig.\ \ref{fig4} for $w<0$.
The corresponding inequivalent sites, for $w>0$, are charted in
Fig.\ \ref{fig5}. We find that there are six inequivalent sites
(defined as those seeing a different vortex environment) in the
Kagome lattice. We now aim to construct different density wave
patterns based on two rules: a) the boson filling must be $f=2/3$
and b) the equivalent sites shown in Fig.\ \ref{fig4} must have the
same filling throughout the lattice\cite{commentden}.

\begin{figure}
\rotatebox{0}{
\includegraphics[width=0.9\linewidth]{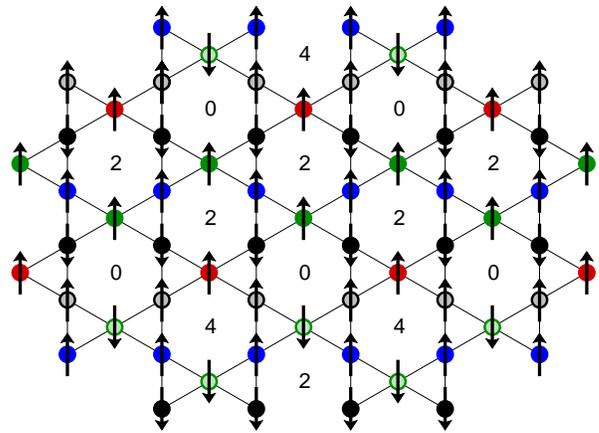}}
\caption{Density wave state $\Phi_2$ at $f=2/3$. Here the hexagons
with black and green (both solid and open circles) dots have net
boson filling $3$ (or magnetization $0$ in spin language). The blue
and red sites are filled (spin up). The centers of the hexagons are
marked with $0$, $2$ and $4$ respectively and reflects the sum of
magnetization of the sites surrounding the hexagons. This state
shows a $9$ by $9$ density wave pattern (full pattern not shown in
the figure). Another state $\Phi_3$ with the same periodicity is
easily obtained from $\Phi_2$ by simply reversing the direction of
the spins (or changing from being filled to being empty and vice
versa) on all the green and the black sites (both open and solid
black and green circles) while leaving the blue and the red sites
unchanged. This has the effect of $2\leftrightarrow 4$ for the
magnetization of the hexagons. Note that any linear superposition
$\Phi(\theta)= \cos(\theta) \Phi_2 + \sin(\theta) \Phi_3$ is also a
possible ground state with $\left<m\right>=1/3$. The state
$\Phi(\pi/4)$ has a periodicity of $3$ by $3$. } \label{fig6}

\end{figure}

Such possible states are sketched in Figs.\ \ref{fig5} and
\ref{fig6}. The state denoted by $\Phi_1$ is shown in Fig.\
\ref{fig5} and has a $3$ by $3$ ordering pattern. Here all the red
and blue sites are empty (or spin down ) while the black and the
green sites (both open and closed circles) are occupied (spin up
sites). The numbers in the center of the hexagons denote the sum of
magnetization (in units of $\hbar/2$) of the sites surrounding the
hexagons. For the state $\Phi_1$ shown in Fig.\ \ref{fig5}, these
take values $0$ and $6$.

More complicated states $\Phi_2$ and $\Phi_3$ with $9$ by $9$
ordering pattern are also possible. These are shown in Fig.\
\ref{fig6}. For the state $\Phi_2$ the bosons are localized in red,
blue, green (closed circle) and black (open circle) sites whereas
the green (open circle) and black (closed circle) sites are vacant.
The net magnetization of hexagons for $\Phi_2$ takes values $0$, $4$
and $2$ as shown in Fig.\ \ref{fig6}. The state $\Phi_3$ can be
similarly obtained from $\Phi_2$ by interchanging the occupations of
the black and green sites while leaving the red and the blue sites
filled. This has the effect of $2\leftrightarrow 4$ for the
magnetization of the hexagons. Interestingly, any linear
combinations of $\Phi_2$ and $\Phi_3$, $\Phi(\theta)= \cos(\theta)
\Phi_2 + \sin(\theta) \Phi_3$ for any arbitrary mixing angle
$\theta$, is also a valid density wave ordered state. The most
interesting among these states turn out to be the $\Phi(\pi/4)$
which has an $3$ by $3$ ordering pattern. Such a state corresponds a
superposition of filled and empty boson sites on the green and black
(both empty and filled) sites whereas the red and blue sites are
filled. Here the sum of boson fillings of the sites surrounding the
hexagons, takes values $0$ and $3$ as can be inferred from Fig.\
\ref{fig6}. Notice that the two states $\Phi_1$ and $\Phi(\pi/4)$
constructed here has the same long-range ordering pattern.

To distinguish between the states $\Phi_1$ and $\Phi(\pi/4)$,
consider the operator ${\mathcal N}_b = \sum_{i \in {\rm hexagon}}
n_i $, where $\left<{\mathcal N}_b\right>$ gives the sum of boson
fillings of the sites surrounding the hexagon with center $b$. The
values of $\left<{\mathcal N}_a\right>$ for different states has
been shown in Figs.\ \ref{fig5} and \ref{fig6}. We note that the
distinction between the states $\Phi_1$ and $\Phi(\pi/4)$ can be
made by computing the values of  $ \left< \sum_a {\mathcal N}_a
{\mathcal N}_{a+u}\right>$ which measures the correlation between $
{\mathcal N}$ operators at sites $a$ and $a+u$ where $u$ is the
basis vector for the dice lattice shown in Fig.\ \ref{fig01}. Deep
inside the Mott phase such correlations should vanish for $\Phi_1$
while it will remain finite for $\Phi(\pi/4)$. Alternatively, one
can also compute the distribution of the hexagons with different
values of $\left<{\mathcal N}_a\right>$ to achieve the same goal
\cite{sergei1}. Such a distribution, computed in Ref.\
\onlinecite{sergei1}, seems to be consistent with the state
$\Phi(\pi/4)$ obtained here.

Recently exact diagonalization study of XXZ model on Kagome lattice
has been carried out in Ref.\ \onlinecite{cabra1}. Their results
concluded the existence of $3$ by $3$ patterned RVB state for
$m=\hbar/6$. We note that the RVB state proposed in Ref.\
\onlinecite{cabra1} shall have identical long range ordering pattern
and $ \left< \sum_a {\mathcal N}_a {\mathcal N}_{a+u}\right>$  to
the state $\Phi(\pi/4)$ obtained in the vortex mean-field theory. Of
course, a true RVB state, if it exists, is beyond the reach of the
mean-field treatment of the vortex theory. Obtaining such states
from a dual vortex analysis is left as an open issue in the present
work.

\section{Discussion}
\label{conclusion}

In this work, we have applied the dual vortex theory developed in
Refs.\ \onlinecite{balents1} to analyze the superfluid and Mott
insulating phases of extended Bose-Hubbard model (Eq.\ \ref{boson1})
(or equivalently XXZ model (Eq.\ \ref{spinmodel1})) on a Kagome
lattice at boson filling $f=1/2, 2/3$. The dual theory developed
explains the persistence of superfluidity in the bosonic model at
$f=1/2$ of arbitrary small values of $t/U, t/V$ seen in the
recent Monte-Carlo studies \cite{sergei1}, and shows that dynamic
localization of vortices due to destructive Aharonov-Bohm
interference, dubbed as ``Aharonov-Bohm caging'' in Ref.\
\onlinecite{vidal1}, is at the heart of this phenomenon. This also
offers an explanation of the absence of $S_z$ ordering in the XXZ
model at zero longitudinal magnetic field for arbitrarily large
$J_z/J_x$ as noted in earlier studies \cite{sondhi1}. In contrast
for $f=2/3$, we find that there is a direct transition from the
superfluid to a translational symmetry broken Mott phase. We have
derived a Landau-Ginzburg action in terms of the dual vortex and
gauge fields to describe this transition. We have shown that the
order and the universality class of the transition depends on
relevance/irrelevance of the sixth order term in the effective
action. In particular, if this term turns out to be irrelevant, the
critical point is deconfined and is accompanied with boson
fractionalization. We have also sketched, within saddle point
approximation of the vortex action, the possible ordered Mott states
that exhibit a $3$ by $3$ ordering pattern. The results obtained
here are in qualitative agreement with earlier numerical studies
\cite{sergei1,sondhi1} on related models.

We thank R. Melko and S. Wessel for helpful discussions and
collaborations on a related project. We are also grateful to
T. Senthil and A. Vishwanath for numerous insightful comments.
This work was supported by the NSERC of Canada, Canada Research Chair
Program, the Canadian Institute for Advanced Research, and
Korea Research Foundation Grant No. KRF-2005-070-C00044.

{\it Noted Added}: While this manuscript is being
prepared, we became aware of a related work by L. Jiang and J. Ye
\cite{jiang1}.

\appendix

\section{Duality transformation}
\label{appc}

In this section, we briefly sketch the duality analysis which leads
to the dual action (Eq.\ \ref{dual}). We start from the Bose-Hubbard
model $H_{\rm boson}$ (Eq.\ \ref{boson1}). First, we follow Ref.\
\onlinecite{steve1}, to obtain an effective rotor model from $H_{\rm
boson}$ given by
\begin{eqnarray}
H_{\rm rotor} &=& -t \sum_{\left<i \alpha\right>}
\cos\left(\Delta_{\alpha} {\hat \phi}_i \right) +  V
\sum_{\left<ij\right>} {\hat n}_i {\hat n}_j \nonumber\\
&& + \sum_i \left[- \mu {\hat n}_i +\frac{U}{2} {\hat
n}_i\left({\hat n}_i-1\right)\right] \label{hrotor}
\end{eqnarray}
where the rotor phase (${\hat \phi}$) and number (${\hat n}$)
operators satisfy the canonical commutation relation $\left[{\hat
\phi}_i,{\hat n}_j \right] = i \delta_{ij}$, $\Delta_{\alpha}$
denotes lattice derivative such that $\Delta_{\alpha} {\hat \phi}_i=
{\hat \phi}_{i+\alpha} - {\hat \phi}_i$ and $i,\alpha=x,y$ runs over
sites of the kagome lattice.

Next, following standard procedures \cite{balents1,steve1}, we write
down the partition function corresponding to $H_{\rm rotor}$ in
terms of path integrals over states at large number of intermediate
time slices separated by width $\Delta \tau$. These intermediate
states use basis of ${\hat n}_i$ and ${\hat \phi}_i$ at alternate
times. The kinetic energy term of $H_{\rm rotor}$ which acts on
eigenstates of ${\hat \phi}_i$ can be evaluated as
\begin{eqnarray}
\exp\left[ t \Delta \tau \cos\left(\Delta_{\alpha} \phi_i \right) \right]\nonumber\\
\rightarrow \sum_{J_{i\alpha}} \exp\left[ -J_{i\alpha}^2/(2 t\Delta
\tau) - i J_{i\alpha} \Delta_{\alpha} \phi_i \right]
\end{eqnarray}
where in the last line we have used the standard Villain
approximation \cite{steve1} and $J_{i\alpha}$ are integer fields
living on the links of the Kagome lattice. Integrating, over the
fields $\phi_i$ and defining the integer field $J_{i \mu} = (n_i,
J_{ix}, J_{iy})$, we get the link-current representation of the
rotor model (Eq.\ \ref{hrotor}) \cite{balents1,steve1}
\begin{eqnarray}
Z &=& \sum_{J_{i\mu}} \exp \Big[ -\frac{1}{2e^2} \sum_{i,\mu}
\left(J_{i \mu} - f \delta_{\mu \tau}\right)^2 \nonumber\\
&& - \Lambda \sum_{\left<ij\right>} J_{i\tau} J_{j\tau} \Big]
\prod_i \delta \left( \Delta_{\mu} J_{i\mu} \right) \label{dac1}
\end{eqnarray}
where $f= \mu/U +1/2$, $\Lambda = V\Delta \tau$, and we have
rescaled the time interval $\Delta \tau$ so that $ e^2 = t\Delta
\tau = 1/U\delta \tau$. The constraint of vanishing divergence of
the currents ($\Delta_{\mu} J_{i\mu}=0$) in the partition function
$Z$ comes from integrating out the phase fields $\phi$. We now note
that this constraint equation can be solved by trading the integer
current fields in favor of gauge fields $A_{b \mu}$ which lives on
the links of the dual(dice) lattice and satisfy $J_{i \mu} =
\epsilon_{\mu \nu \lambda} \Delta_{\nu} A_{b \lambda}$. Note that
$\epsilon_{\tau \nu \lambda}\Delta_{\nu} A_{b \lambda}$ corresponds
to the physical boson density $n_i \equiv J_{i \tau}$. In terms of
these fields the partition function becomes
\begin{eqnarray}
Z &=&  \sum_{A_{b\mu}}\exp \left[-\frac{1}{2e^2} \sum_b
\left(\epsilon_{\mu \nu \lambda} \Delta_{\nu} A_{b\lambda} -
f\delta_{\mu \tau} \right)^2 \right] \label{dac2}
\end{eqnarray}
Here we have dropped the term proportional to $\Lambda$ and have
assumed that it's main role is to renormalize the coefficient $e^2$
\cite{balents1}. Next, we promote the integer-valued fields $A_{b
\mu}$ to real valued fields by using Poisson summation formula and
soften the resulting integer constraint by introducing a fugacity
term $y_v \cos\left(2\pi A_{b \mu}\right)$. Further, we make the
$U(1)$ gauge structure of the theory explicit by introducing the
rotor fields $\theta_b$ on sites of the dual lattice, and mapping
$2\pi A_{b\mu} \rightarrow 2\pi A_{b\mu} - \Delta_{\mu} \theta_b$.
This yields
\begin{eqnarray}
Z &=& \int {\mathcal D} A \int {\mathcal D}\theta \exp
\Bigg[ \nonumber\\
&& -\frac{1}{2e^2} \sum_b \left(\epsilon_{\mu \nu \lambda}
\Delta_{\nu} A_{b\lambda} - f\delta_{\mu \tau} \right)^2 \nonumber\\
&& + y_v \sum_b \cos\left( 2\pi A_{b \mu} - \Delta_{\mu} \theta_b
\right) \Bigg] \label{dac3}
\end{eqnarray}
We note that $ \exp(i\theta_b)$ corresponds to the vortex creation
operator for the original bosons \cite{balents1}. Finally, we trade
off $\exp\left(i \theta_b \right)$ in favor of a ``soft-spin'' boson
field $\psi_b$ as in Ref.\ \onlinecite{balents1} to obtain the final
form of the effective action
\begin{eqnarray}
Z &=& \int {\mathcal D} A \int {\mathcal D}\theta
 \exp\left(-S_d\right), \nonumber\\
S_d &=&  \frac{1}{2e^2} \sum_b \left(\epsilon_{\mu \nu \lambda}
\Delta_{\nu} A_{b\lambda} - f\delta_{\mu \tau} \right)^2 \nonumber\\
&& - y_v \sum_b \left(\psi_{b +\mu} e^{2\pi i A_{b \mu}} \psi_b +
{\rm h.c.} \right) \nonumber\\
&&  + \sum_b \left( r \left|\psi_b \right|^2 + u \left|\psi_b
\right|^4 \right)
\end{eqnarray}

\section{ Transformation of $\Psi$ }
\label{appa}

Here we present the details of derivation of the transformation
properties of $\psi$ under rotation by $ \pi/3$. The rest of the
transformation properties can be derived in a similar way.

First note that any general term in the Hamiltonian $H_{\rm
kinetic}$ (Eq.\ \ref{kinetic}) can be written as
\begin{eqnarray}
\sum_{{\bf a}^{(1)},{\bf a}^{(2)}} \psi_{\alpha}^{\ast}
\left(a_x^{(1)},a_y^{(1)}\right) \psi_{\beta
}\left(a_x^{(2)},a_y^{(2)}\right) \exp\left[i \theta({\bf
a}^{(1)},{\bf a}^{(2)})\right] \nonumber\\ \label{b1a}
\end{eqnarray}
where
\begin{eqnarray}
 \theta({\bf a}^{(1)},{\bf a}^{(2)})&=& \frac{2\pi f}{\sqrt{3}}
 \left(a_x^{(1)}+a_x^{(2)}\right)\left(a_y^{(2)}-a_y^{(2)}\right). \label{b1b}
\end{eqnarray}
Here $\alpha,\beta$ are site index which can take values $A$,$B$ or
$C$, ${\bf a}^{(2)}$ denote a near neighbor site of ${\bf a}^{(1)}$,
and the phase factor $ \theta = \frac{2\pi}{\phi_0}
\int_{a_y^{(1)}}^{a_y^{(2)}} A_y dy $ is obtained with the gauge
$A=A_y= Hx$ and $f=\phi/\phi_0$ where $\phi = H a^2 \sqrt{3}/2$ is
the dual flux through an elementary rhombus of the dice lattice
\cite{vidal1} and $\phi_0$ is the flux quanta. For example a typical
such term can be
\begin{eqnarray}
\sum_{{\bf a}^{(1)}} \psi_{A}^{\ast}\left(a_x^{(1)},a_y^{(1)}\right)
\psi_{C}\left(a_x^{(1)}+1/2,a_y^{(1)}+\sqrt{3}/2\right)\nonumber\\
 \times e^{2\pi i f \left(a_x^{(1)}+1/4 \right)} \label{b2}
\end{eqnarray}
Now consider a rotation by an angle $\pi/3$. After the rotation, a
typical term in the Hamiltonian becomes
\begin{eqnarray}
\sum_{{\bf a}^{(1)},{\bf a}^{(2)}} \psi_{\alpha}^{'\ast}
\left(a_x^{'(1)},a_y^{'(1)}\right) \psi'_{\beta
}\left(a_x^{'(2)},a_y^{'(2)}\right)\nonumber\\
 \times e^{2\pi i f
\left(a_x^{'(1)}+a_x^{'(2)}\right)\left(a_y^{'(2)}-a_y^{'(2)}\right)
/\sqrt{3}} \label{b3}
\end{eqnarray}
where the rotated coordinates are given in terms of the old
coordinates by
\begin{eqnarray}
a'_x &=& a_x/2 -\sqrt{3} a_y/2 \nonumber\\
a'_y &=& a_y/2 +\sqrt{3} a_x /2 \label{b4}
\end{eqnarray}
Now from the structure of the dice lattice from Fig.\ \ref{fig1}, we
know that such a rotation interchanges $B$ and $C$ sites while
transforming $A$ sites onto themselves. Then comparing terms
\ref{b1a} and \ref{b3}, we see that we need
\begin{eqnarray}
\psi'_{\beta} \left(a'_x,a'_y\right) e^{- i \alpha(a'_x,a'_y)} &=&
\psi_{\beta'}(a_x,a_y) \label{b5}
\end{eqnarray}
where $\beta$ takes values $A$,$C$ and $B$ for $\beta'=A,B,C$.
Therefore we have
\begin{eqnarray}
\alpha\left({\bf a}^{'(1)}\right)-\alpha\left({\bf a}^{'(2)}\right)
&=& \theta({\bf a}^{(1)},{\bf a}^{(2)})-\theta({\bf a'}^{(1)},{\bf
a'}^{(2)}) \nonumber\\ \label{b6}
\end{eqnarray}
Fortunately one has a relatively straightforward solution to Eq.\
\ref {b6}. Rewriting ${\bf a}^{(1)}$ and $ {\bf a}^{(2)}$ in terms
of ${\bf a'}^{(1)}$ and $ {\bf a'}^{(2)}$
\begin{eqnarray}
a_x &=& a'_x/2 +\sqrt{3} a'_y/2 \nonumber\\
a_y &=& a'_y/2 -\sqrt{3} a_x /2 \label{b7}
\end{eqnarray}
and using Eq.\ \ref{b1b}, we get, after some algebra
\begin{eqnarray}
\alpha(a'_x,a'_y) = - 2 \pi i f \left[ \frac{1}{4} \left(a_x^{'2} -
a_y^{'2} \right) - \sqrt{3} a'_x a'_y /2 \right] \label{b8}
\end{eqnarray}
Using Eqs.\ \ref{b8} and \ref{b5}, one obtains
\begin{eqnarray}
\psi'_{\beta}(a_x,a_y) &=& e^{i\alpha(a_x,a_y)}
\psi_{\beta'}(a"_x,a"_y)
\end{eqnarray}
where $ a"_x = (a_x + \sqrt{3}a_y)/2$ and $ a"_y = (a_y -
\sqrt{3}a_x)/2$. This is Eq.\ \ref{rot2} of the main text where we
have used $a_x \equiv x$ and $a_y \equiv y$ for notational brevity.
All other transformation properties can be obtained in a similar
manner.
\\

\section{Transformation of vortex fields $\varphi$}

\label{appb}

Here we consider the transformation of $\varphi_{1,2}$. First let us
concentrate on the translation operator $T_u$. Under action of
$T_u$, $\Phi \rightarrow \Phi'= \Phi(x-u_x,y-u_y) \exp\left(4\pi i f
u_x y/\sqrt{3}\right)$. Note that for $f=2/3$, $u_x=3a/2$ and
$y=\sqrt{3}a n /2$, the exponential factor become unity. Thus one
gets
\begin{eqnarray}
\Phi' &=& \exp\left[i\pi(n-1)/3\right] (c,-c,0) \varphi_1 \nonumber\\
&& + \exp\left[i\pi (m-1) +2i\pi(n-1)/3 \right] (c,0,-c) \varphi_2
\nonumber\\ \label{t1}
\end{eqnarray}
Comparing Eq.\ \ref{t1} with Eqs.\ \ref{efunc} and \ref{vfunc}, one
gets
\begin{eqnarray}
\varphi_1 \rightarrow \varphi_1 \exp(-i\pi/3) \nonumber\\
\varphi_2 \rightarrow -\varphi_2 \exp(-2\pi i/3)
\end{eqnarray}
Similarly we get the transformation properties of $\varphi_{1,2}$
under $T_v$ and $I_x$.

The transformation under rotation $\pi/3$ is slightly more
complicated. First we need to note that for $x=3ma/2$ and
$y=\sqrt{3}na/2$, one has
\begin{eqnarray}
\omega^{\left[\left(y^2-x^2\right)/4 + \sqrt{3}x y/2\right]} &=&
\exp \left[ -3\pi i (n-m)^2/4 + \pi i n^2 \right] \nonumber\\
\end{eqnarray}
Next note that with our choice of origin all the A sites are such
that $m+n$ is an even integer and in such cases
\begin{eqnarray}
\exp \left[ -3\pi i (n-m)^2/4 + \pi i n^2 \right] &=& \exp \left[
\pi i (m-3n)/2 \right] \nonumber\\
\end{eqnarray}
for all integers $m$ and $n$. Hence under a $\pi/3$ rotation, we
have using Eq.\ \ref{rot2}
\begin{eqnarray}
\Phi' &=& \Big[\exp\left[-\pi i(m+3n)/6\right] (c,0,-c) \varphi_1 \nonumber\\
&& \exp\left[-3\pi i m/2 +\pi i n/6 \right] (c,-c,0) \varphi_2 \Big]
\nonumber \\
&& \times \exp \left[- \pi i (m-3n)/2 \right] \nonumber\\
&=& \exp[i\pi (m+2n/3)] (c,0,-c) \, \varphi_1 \nonumber\\
&& + \exp[i\pi n/3] (c,-c,0) \, \varphi_2  \label{t2}
\end{eqnarray}
Comparing Eq.\ \ref{t2} with Eqs.\ \ref{efunc} and \ref{vfunc}, one
gets $\varphi_{1(2)} \rightarrow \varphi_{2(1)}$. A similar analysis
yields $\varphi_{1(2)} \rightarrow \varphi_{1(2)}$ under the
operation $R_{2\pi/3}$.


\begin{thebibliography}{99}
\bibitem{comment1} A milder version of such a situation may occur
for square lattice for large $q$ since the depth of the vortex
minima and the width of the lowest Hopfstder band become smaller
with increasing $q$ as pointed out in Ref.\ \onlinecite{balents1}.
However, the phenomenon seen here at $q=2$ has a different physical
origin and is much more robust.
\bibitem{commentden} A more formal way of doing this would be to
construct the density operators as bilinears of the vortex fields
with appropriate transformation properties \cite{balents1,burkov1}.
We have not attempted this in the present work.
\end{thebibliography}
\end{document}